\documentclass[oneside,11pt]{article}
\usepackage{amsmath,amsfonts,amsthm,amssymb,epsfig}
\newcommand{\pr}[1]{\left( #1\right)}
\newcommand{\prr}[1]{\left[ #1 \right]}
\newcommand{\es}[1]{\begin{equation}\begin{split}#1\end{split}\end{equation}}

\newcommand{\R}{\mathbb{R}}
\newcommand{\N}{\mathbb{N}}
\small\normalsize

\begin{document}
\title{Full Connectivity: Corners, edges and faces}
\author{Justin Coon,$^{1}$ Carl P. Dettmann and Orestis Georgiou$^{3^\ast}$\\
\\
\normalsize{$^{1}$ Toshiba Telecommunications Research Laboratory}\\
\normalsize{32 Queen Square, Bristol BS1 4ND, United Kingdom}\\
\normalsize{$^{2}$ University of Bristol School of Mathematics}\\
\normalsize{University Walk, Bristol BS8 1TW, United Kingdom}\\
\normalsize{$^{3}$ Max-Planck-Institute for the Physics of Complex Systems}\\
\normalsize{N\"{o}thnitzer Str. 38, 01187 Dresden, Germany}\\
\\
\normalsize{$^\ast$To whom correspondence should be addressed.}}
\date{}
\maketitle

\begin{abstract}
We develop a cluster expansion for the probability of {\em full connectivity} of high
density random networks in confined geometries. In contrast to
percolation phenomena at lower densities, boundary effects, which
have previously been largely neglected, are not only relevant but dominant.
We derive general analytical formulas that show a persistence of
universality in a different form to percolation theory, and provide
numerical confirmation. We also demonstrate the simplicity of our approach in three simple
but instructive examples and discuss the practical benefits of its application to different models.
\end{abstract}

\section{Introduction}

Percolation is a phase-transition phenomenon in large random networks whereby at a critical value
$\rho_c$ of a parameter such as density that controls the connection probabilities,
the largest connected component (cluster) of the system
experiences a sudden change from being independent of system size (microscopic) to being proportional to it (macroscopic).
For example in infinite systems with densities $\rho<\rho_{c}$ (sub-critical), all clusters are finite
almost surely while for $\rho>\rho_{c}$ (super-critical), any given node is in an infinite cluster with
positive probability. As with other statistical phase transitions, in the thermodynamic limit of
large system size, the percolation density $\rho_{c}$ is
very much independent of the size and shape of the system and in general of the
microscopic details of the model leading to the phenomenon of universality. Percolation theory does
not, however, address the question of when finite systems are fully connected, that is, the
probability of the entire system comprising a single cluster. Here we develop a theory of the
latter, showing that boundary effects are crucial for understanding this problem, and demonstrate
a general formalism for calculating their contributions.

The well developed theory of percolation \cite{Bela06} finds its historic roots in the early 1930s in studies by physicists and chemists concerned with nucleation and condensation of gases into liquids as well as clustering of interacting particles and colloids. The theory of percolation was then initiated in 1957
in order to study random physical processes such as fluid flow through disordered porous media \cite{Hammersley57}.
A more general mathematical approach was formulated soon thereafter in 1959 in the form of random graphs \cite{Erdos59} using the famous Kolmogorov zero-one law in probability theory to state that given a infinite graph with connections chosen randomly and independently, the existence probability of an infinite cluster is either zero or one. A wealth of different models and approaches as well as outstanding open problems can be found in \cite{Bela01}. The exact statistical mechanical formalism of continuum percolation on a lattice-free basis was eventually formulated in the late 1970s \cite{Coniglio77} and has been extensively applied in a wide variety of settings as the theory provides useful information on cluster statistics (for a review article see \cite{Stell96}).

The triumph of percolation theory and statistical physics \cite{Hill56} in describing particle clustering in liquids \cite{DeSimone86}, gases \cite{Hill55,Chiew83} and colloids \cite{Safran85}, was also successfully adopted in studies of electrical conductance in disordered media. These have included investigations of transport in carbon nano-tube networks \cite{Hu04} and metallic insulation in composite materials \cite{Toker03}. The concept of quantum percolation and its connections with the quantum Hall effect \cite{Wang93} as well as phenomena such as spontaneous magnetization \cite{Kaminski02} are but a few of the many recent theoretical advancements.
Moreover, percolation theory has gone beyond physical sciences to topics such as network modeling. These include interpreting (self)-organization processes of complex networks, such as those displayed by functionally related proteins \cite{Kim02}; the spread of forest fires \cite{Drossel92,Pueyo10}, epidemics \cite{Moore00,Miller09,Danon11}, computer \cite{Cohen03} and phone \cite{Wang09} viruses; industrial and economic sectors \cite{Duffie07}; and social groups of people \cite{Palla05,Palla07,Parshani11}.

\begin{figure}[t]
\begin{center}
\includegraphics[scale=0.2]{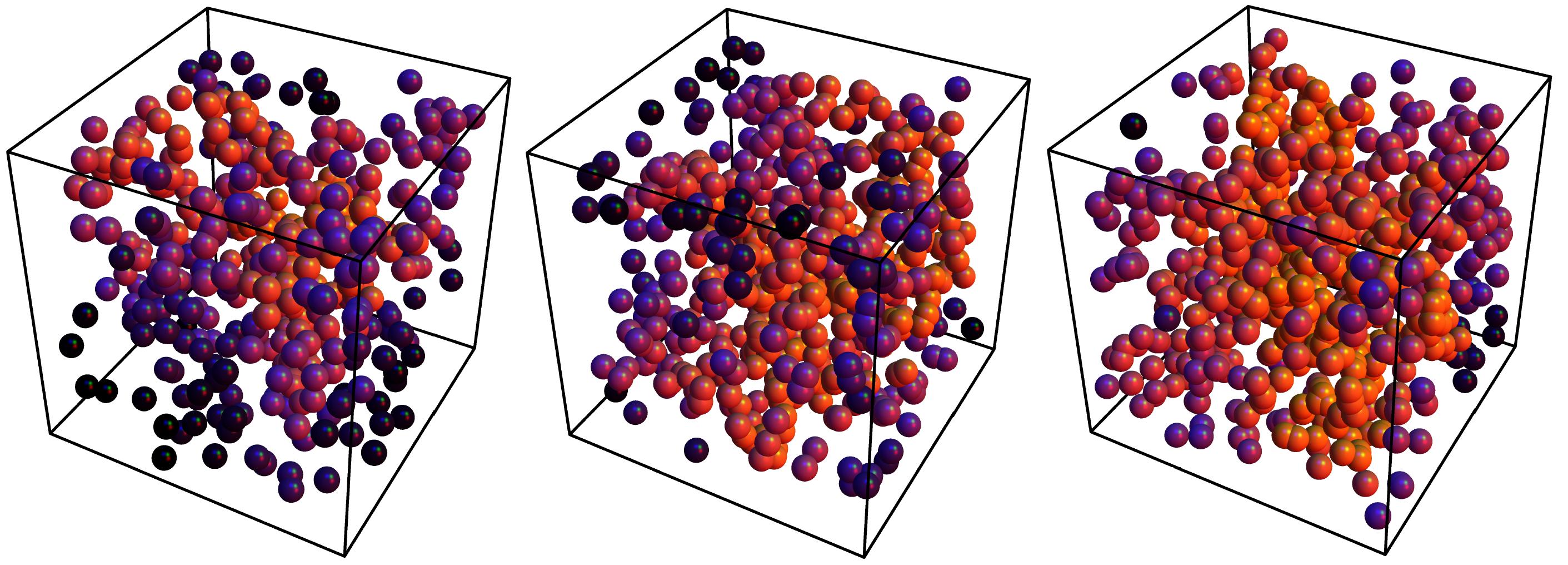}
\caption{\label{fig:balls} \footnotesize (Color online) A single high density realization is shown with $N=450, 500$ and $600$ nodes (represented as balls) randomly placed inside a cubed domain of side $L=10$. Lighter colors indicate a higher probability of being in the largest connected component; note that isolated nodes (darker colors) are concentrated at the edges and corners at higher densities.}
\end{center}
\end{figure}
The current work is primarily motivated by the recent adaptation of percolation theory in wireless communications and multihop relay networks. These consist of a number of communication devices (nodes) which can pass messages to each other without the need of a central router.
Multihop relay networks can therefore achieve good and reliable coverage and connectivity over a large area
even when some nodes are moved or deactivated.
Most models of such systems are closely related or derived from studies by probabilists under the topic of random geometric graphs or networks.
Percolation theory, both on a lattice and continuum has thus been previously applied to random networks by engineers identify and analyze power management techniques \cite{Glauche03,Paschos09}, network resilience \cite{Callaway00}, efficient relay placement \cite{Li09}, coverage and connectivity in wireless sensor networks \cite{Ammari08} (a direct adaptation of \cite{Chiew83}), and the information theoretic capacity of networks \cite{Franceschetti07}.

A major weakness of the theory which is immediately evident in the current context is that many networks, although large, are not of infinite size and are often confined within a finite region. Therefore infinite (or `giant') clusters and critical percolation densities no longer make sense in the usual thermodynamic limit. Instead, the more suited notion of full connectivity (originally posed in \cite{Erdos59}) is addressed, typically in some asymptotic regime such as in the extended network and the dense network models (see \cite{Mao11} and references therein). Both of these models scale the system size exponentially with density and hence manage to ignore any boundary effects.  However in the natural limit that we consider, where system size increases with density more slowly than exponential
or not at all, boundary effects dominate.

\begin{figure}[t]
\begin{center}
\includegraphics[scale=0.25]{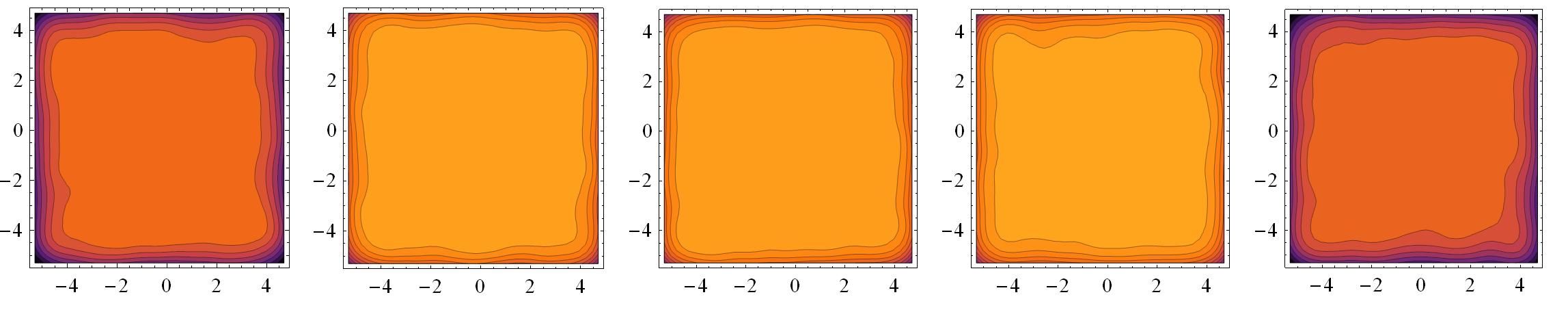}
\caption{\label{fig:cross} \footnotesize (Color online) Five equally spaced cross sections of the average connectivity (averaged over $100$ different realizations) inside a cubed domain of side $L=10$ with node density $\rho=0.5$. The outer plots correspond to opposite faces of the cube. The same color scheme as in Figure~\ref{fig:balls} is used with light and dark colors indicating good and bad connectivity respectively. It is clear that corners, edges and faces of the cube constitute critical regions where full connectivity is lost with maximum probability.}
\end{center}
\end{figure}
The above observation is emphasized in Figure~\ref{fig:balls}, which shows a single realization of (from left to right), $450$, $500$ and $600$ nodes (represented as overlapping balls) placed randomly inside a cubed domain, with each ball colored according to the total probability of it connecting to any of the other $N-1$ nodes surrounding it. We assume here an exponentially decaying (with distance) connection probability function (see equation~\eqref{H}). Notice that isolated, and hence hard to connect, nodes (black) are in this case near corners, edges or faces of the cubed domain. Averaging over 100 such realizations and producing density plots of cross-sections of the cubed domain (see Figure~\ref{fig:cross}) clearly demonstrates that these are critical regions where full connectivity is lost with maximum probability.
That is, to leading order in the high density limit, the probability of obtaining a fully connected network is controlled by the probability of a single isolated node. Lower density corrections to this are due to two, three and larger isolated clusters. This observation has formed the basis of our recent work \cite{CDG11PRE}, where we presented a new cluster-expansion approach to analyzing connectivity in confined geometries. This approach allowed for a consistent analysis and inclusion of the local geometric boundary effects leading to a general closed-form analytical formula for calculating the probability of full connectivity $P_{fc}$.

The current paper complements our previous results by providing additional details of our approach in a more general setting, extends them with further calculations of second order correction terms, and also demonstrates the inaccuracy of conventional bulk models through three simple, yet instructive example network confinements (circle, square and sector) which are numerically confirmed for both probabilistic and sharp (unit-disk type) pair connectedness functions.
The structure of the paper is as follows: In section 2, we derive a high density cluster approach which we then expand up to second order in subsections 2.1, 2.2 and 2.3 assuming a homogeneous system. We then compute the overall probability of a fully connected network for a given exponentially decaying connectivity function in subsection 2.4. In section 3, we lift the homogeneity assumption and consider boundary effects explicitly illustrating their importance for any connectivity probability function that decays suitably fast. We then work through three examples (subsections 3.2, 3.3 and 3.4) for the previously given connectivity function and provide plots of comparisons with numerical simulations. These examples lead to the construction of a general formula for the overall probability of a fully connected network in subsection 3.5. Finally in section 4 we summarize and conclude with a short discussion.

\section{A cluster expansion approach}

In this section we develop a high density connectivity model based on a cluster expansion approach which is able to produce closed-form analytical results in some generality. We assume a homogeneous network and perform the expansion up to second order and comment on the third and higher order corrections. First however we need to set the stage and define some useful quantities.

Consider $N$ randomly distributed nodes with locations ${\bf r}_i\in{\cal
V}$, a convex subset of $\mathbb{R}^d$, where $i=1,2,\ldots N$, according to a uniform
density $\rho=N/V$, where $V=|{\cal V}|$ and
$|\cdot|$ denotes the size of the set using the Lebesgue measure of the
appropriate dimension or the cardinality of a finite set. We assume that nodes
$i$ and $j$ are directly connected with probability $H(r_{ij})$, often written just $H_{ij}$ where
$r_{ij}=|{\bf r}_j-{\bf r}_i|$ is the distance between $i$ and $j$.

We may also consider a more general setting, as follows:
$\mathbb{R}^d$ could be replaced by a Riemannian manifold ${\cal M}^d$
with volume element derived from its metric tensor. The
direct connection distance function $r_{ij}$ may or may not be given by
geodesic distance derived
from the metric tensor.  The convexity assumption is then replaced by
the statement that this distance function is a metric in the mathematical
sense, \textit{i.e.} symmetric,
non-negative, zero only when the nodes coincide, and satisfying the
triangle inequality.\footnote{
If the convexity assumption is relaxed it is a semi-metric function, \textit{i.e.}
the triangle inequality may not be satisfied. For example, $r_{ij}$
is effectively infinite if there is an obstacle blocking the path
of a wireless signal, but adding a node $k$ that avoids the obstacle may permit
indirect connection between $i$ and $j$, thus in this case
$r_{ij}>r_{ik}+r_{kj}$.  We defer the treatment of obstacles to a
future paper.}
One example of a non-Euclidean manifold and
discussion of its curvature effects is discussed in section 2.4.
Otherwise we will use only Euclidean space $\mathbb{R}^d$.

We define an average over all configurations
\es{\langle A\rangle = \frac{1}{V^{N}}\int_{\mathcal{V}^{N}} A(\mathbf{r}_{1},\mathbf{r}_{2},\ldots, \mathbf{r}_{N}) \mathrm{ d}\mathbf{r}_{1} \mathrm{ d}\mathbf{r}_{2}\ldots \mathrm{ d}\mathbf{r}_{N}
.}
We also need notation to define the relevant graphs.  Let $S=\{1,2,3,\ldots,N\}$.
A graph $g=(A,L)$ consists of a set $A\subseteq S$ of nodes, together
with a collection $L\subseteq \{(i,j)\in A:i<j\}$
of direct links, that is unordered distinct pairs of nodes. As a slight abuse of notation we write
$(i,j)\in g$ to denote that $(i,j)$ is an element of the set of links $L$ associated with the graph $g$.
We write $G^A$ for the set of graphs with nodes in $A$, and $G^A_j$ for the set with nodes in
$A$ and largest connected component (cluster) of size $j$ with $1\leq j\leq |A|$.

The probability that two nodes are connected or not leads to the trivial identity:
\es{\label{1}1\equiv H_{ij}+ (1-H_{ij}).}
Multiplying over all links with nodes in a set $A$ expresses the probability of all possible
combinations. This can be written as
\es{\label{2}1= \prod_{i,j\in A; i<j}[H_{ij}+ (1-H_{ij})]= \sum_{g\in G^A}\mathcal{H}_{g},}
where
\es{\mathcal{H}_{g}=\prod_{(i,j)\in g}H_{ij}\prod_{(i,j)\not\in g}(1-H_{ij}).}
The sum in equation~\eqref{2} contains $2^{|A|(|A|-1)/2}$ separate terms. Setting $A=S$, this can be expressed as collections of terms determined by their largest cluster:
\es{\label{3} 1=\underbrace{\sum_{g\in G^{S}_N}\mathcal{H}_{g}}_{P_{fc}(\mathbf{r}_{1},\ldots \mathbf{r}_{N})}+ \sum_{g\in G^{S}_{N-1}}\mathcal{H}_{g} + \ldots +\underbrace{\sum_{g\in G^{S}_{1}}\mathcal{H}_{g}}_{\prod_{i<j}(1-H_{ij})} .}

For a given configuration of node positions $\mathbf{r}_{i}\in\mathcal{V}$, assuming that the nodes are pairwise connected with independent probabilities $H_{ij}$, the first term in equation~\eqref{3} is the probability of being fully connected $P_{fc}(\mathbf{r}_{1},\ldots \mathbf{r}_{N})$. The average of this quantity over all possible configurations $P_{fc}=\langle  P_{fc}(\mathbf{r}_{1},\ldots \mathbf{r}_{N}) \rangle$ is the overall probability of obtaining a fully connected network and is our desired quantity of interest. Hence, rearranging equation~\eqref{3} allows us to obtain expressions for $P_{fc}$ in a consistent way while keeping track of correction terms.

\subsection{Zeroth order approximation}

In the very high density limit of $\rho\rightarrow\infty$, the right hand side of equation~\eqref{3} is dominated by the first term
\es{P_{fc} \approx 1,}
and hence the network is fully connected with probability one. The approximation symbol is used here and from now on to indicate that higher order terms are being ignored.

\subsection{First order approximation}

The first order approximation is obtained when the second term in equation~\eqref{3} is expanded out explicitly. This takes into account all the ways of having an $N-1$ cluster of nodes. Thus the overall probability of a fully connected network is
\es{\label{4} P_{fc} & \approx 1- \langle \sum_{g\in G^{S}_{N-1}}\mathcal{H}_{g}\rangle \\
&= 1- \langle\pr{\sum_{\wp=1}^{N}\prod_{j\neq\wp}(1-H_{j\wp})}\underbrace{\pr{\sum_{g\in G^{S\backslash\{\wp\}}_{N-1}}\mathcal{H}_{g}}}_{\approx1}\rangle \\
& = 1- N\langle\prod_{j=1}^{N-1}(1-H_{j N})\rangle \\
& = 1- \frac{N}{V^{N}} \int_{\mathcal{V}^{N}} \prod_{j=1}^{N-1}(1-H(\mathbf{r}_{jN})) \mathrm{ d}\mathbf{r}_{1}\ldots \mathrm{ d}\mathbf{r}_{N}\\
&=1-\frac{N}{V}\int_{\mathcal{V}}\left(1-\frac{1}{V}\int_{\mathcal{V}}
H({\mathbf r}_{1N})\mathrm{d}{\mathbf r}_1\right)^{N-1} \mathrm{d}\mathbf{r}_N,}
since all nodes are identical and therefore the sum over $\wp$ can be factored out.

Assuming that the network is homogeneous (\textit{i.e.} there are no boundary effects and therefore the system is symmetric under translational transformations) allows for the change in variables
$\mathbf{r}=\mathbf{r}_{1}-\mathbf{r}_{N}$ so the integrals decouple and we get
\es{\label{6} P_{fc} &\approx 1-
N \pr{1-\frac{1}{V}\int_{\mathcal{V}} H(\mathbf{r}) \mathrm{ d}\mathbf{r}}^{N-1}\\
&= 1- N e^{-\rho\int_{\mathcal{V}} H(\mathbf{r}) \mathrm{ d}\mathbf{r}}\prr{1  + \frac{1}{N}\pr{\rho\int_{\mathcal{V}} H(\mathbf{r}) \mathrm{ d}\mathbf{r} - \frac{\pr{\rho \int_{\mathcal{V}} H(\mathbf{r}) \mathrm{ d}\mathbf{r}}^{2}}{2} } + \mathcal{O}\pr{\frac{\rho^{4}}{N^{2}}} },
}
for $N$ large.

At this point we are in a position to discuss the various scaling limits and approximations appearing
in the theory. Until now we have assumed only that the probability of full connectivity is high, which
is reasonable for typical applications such as wireless networks.
This means that $\rho V_H$ is large where $V_H=\int_{\mathcal{V}}H({\mathbf r})\mathrm{d}\mathbf{r}$
is the effective connectivity volume associated with $H(r)$.
Increasing $N$ and $V$ at constant $\rho$ will however
decrease $P_{fc}$  due to the factor of $N$ in front of the exponential.  We see that at fixed $P_{fc}$,
system size scales exponentially with density.  While this is a popular scaling in the literature,
it has the effect of hiding boundary effects, which we will see
are very important. Alternatively we can increase $\rho$ while keeping $V$ fixed.
For this purpose, and for any scaling in which $V/\rho$ decreases, we must use the first line in equation~\eqref{6} as the second line clearly does not converge. However it is also advantageous to
assume $V/V_H$ is large, for example to separate effects due to the range of direct connections
which is proportional to $V_H^{1/d}$, and effects due to the finite extent of the whole system.  In this
case, so for any scaling where $V$ is between linear and exponential in $\rho$, we can use
the simpler second line of equation~\eqref{6}. We will take the latter approach unless otherwise stated.

\subsection{Second order approximation}

The second order approximation involves enumerating the $N-2$ cluster terms, together with the first approximation of the $G^{S\backslash\{\wp\}}_{N-1}$ term in equation~\eqref{4} such that
\es{\label{8} P_{fc}  \approx 1 -\langle\sum_{g\in G^S_{N-1}}\mathcal{H}_{g}\rangle -\langle\sum_{g\in G^S_{N-2}}\mathcal{H}_{g}\rangle.
}
We first examine the first approximation of the $G^S_{N-1}$ term thus including all possible ways of getting an $N-2$ cluster in a $N-1$ node network:
\es{\label{8a}\langle\sum_{g\in G^S_{N-1}}\mathcal{H}_{g}\rangle = \langle \sum_{\wp=1}^{N}\prod_{j\neq\wp}(1-H_{j\wp})\underbrace{\sum_{g\in G_{N-1}^{S\backslash\{\wp\}}}\mathcal{H}_{g}}_{\approx 1- \sum_{g\in G_{N-2}^{S\backslash\{\wp\}} }\mathcal{H}_{g}} \rangle
,}
and
\es{\langle \sum_{g\in G_{N-2}^{S\backslash\{\wp\}} }\mathcal{H}_{g}\rangle = \frac{1}{2}\langle \sum_{\ell=1}^{N-1}\prod_{j\neq\ell}(1-H_{j\ell})\underbrace{\sum_{g\in G^{S\backslash\{\wp,\ell\}}_{N-2}}\mathcal{H}_{g}}_{\approx 1}\rangle
.}
In this way we avoid double counting of different decompositions of subgraphs.
Expanding equation~\eqref{8a} out we have
\es{\label{9} \langle \sum_{g\in G^S_{N-1}}\mathcal{H}_{g}\rangle &= \langle \sum_{\wp=1}^{N}\prod_{j\neq\wp}(1-H_{j\wp}) - \frac{1}{2}\pr{\sum_{\wp=1}^{N} \overbrace{\prod_{j\neq\wp}(1-H_{j\wp})}^{(N-1) \mathrm{ terms}} \sum_{\ell\neq\wp}\overbrace{\prod_{j\neq\ell,\wp}(1-H_{j\ell})}^{(N-2) \mathrm{ terms}}} \rangle \\
&= N \langle \prod_{j=2}^{N}(1-H_{1j}) - \frac{N(N-1)}{2}\pr{(1-H_{12}) \prod_{j=3}^{N}(1-H_{1j})(1-H_{2j}) } \rangle,}
since all nodes are identical.

We now consider the last term in~\eqref{8}.
There are two possible ways of getting an $N-2$ cluster. These correspond to having two isolated nodes and one large $N-2$ cluster, or having a small $2$-node cluster and a large $N-2$ cluster. Hence, adding these two (disjoint) possibilities together we have that
\es{\label{11b} \langle\sum_{g\in G^S_{N-2}}\mathcal{H}_{g}\rangle &= \langle \pr{\sum_{\wp<\ell}\prod_{j\neq \wp}(1-H_{j\wp}) \prod_{\substack{j\neq \wp,\ell}}(1-H_{j\ell}) } \underbrace{\pr{\sum_{g\in G^{S\backslash\{\wp,\ell\}}_{N-2}}\mathcal{H}_{g}}}_{\approx 1} \rangle \\
&+ \langle \pr{\sum_{\wp<\ell}H_{\wp\ell}\prod_{\substack{j\neq \wp,\ell}}(1-H_{j\wp}) (1-H_{j\ell}) } \underbrace{\pr{\sum_{g\in G^{S\backslash\{\wp,\ell\}}_{N-2}}\mathcal{H}_{g}}}_{\approx 1} \rangle,\\
&= \frac{N(N-1)}{2} \langle (1-H_{12})\prod_{j=3}^{N}(1-H_{1j})(1-H_{2j}) \rangle \\
&+ \frac{N(N-1)}{2}\langle H_{12} \prod_{j=3}^{N}(1-H_{1j})(1-H_{2j}) \rangle, \\
&= \frac{N(N-1)}{2}\langle \prod_{j=3}^{N}(1-H_{1j})(1-H_{2j}) \rangle
,}
due to the identity of equation~\eqref{1}.

Putting it together we find a clear sum of zeroth, first and second order terms:
\es{\label{11c} P_{fc}\approx 1-N\langle\prod_{j=2}^N(1-H_{1j})\rangle
-\frac{N(N-1)}{2}\langle H_{12}\prod_{j=3}^N(1-H_{1j})(1-H_{2j})\rangle}
This is our most general result, allowing any dimension, region, and suitably decaying function
$H(r)$.  Assuming that the network is homogeneous as before gives
\es{\label{11d} P_{fc}\approx 1
-N\left(1-\frac{1}{V}\int_{\mathcal{V}}H(\mathbf{r})\mathrm{d}\mathbf{r}\right)^{N-1}
-\frac{N(N-1)}{2V}\int H(\mathbf{r})\left(1-\frac{K(\mathbf{r})}{V}\right)^{N-2}\mathrm{d}\mathbf{r}}
where
\es{\label{10}K(\mathbf{r}_1)=V-\int_{\mathcal{V}}(1-H(\mathbf{r}_2))(1-H(\mathbf{r}_{12}))
\mathrm{d}\mathbf{r}_2}
is a quantity that remains bounded as $V$ increases. For sufficiently large $V$ (see \eqref{6} above)
we can approximate this as
\es{\label{13}P_{fc}
&\approx 1- N e^{-\rho\int_{\mathcal{V}} H(\mathbf{r}) \mathrm{ d}\mathbf{r}} - \frac{\rho N}{2} \int_{\mathcal{V}} H(\mathbf{r}) e^{-\rho K(\mathbf{r})}  \mathrm{ d}\mathbf{r}.}

\subsection{The $H(\mathbf{r})$ function and connectivity on a sphere}

Translational invariance is unlikely to be a good assumption in practice, except if the nodes are situated on the surface of some $n$-sphere $S^{n}\subset\R^{n+1}$ with $n\geq1$ such that the system is symmetric under rotations about the origin. Here, we consider the realistic case of $n=2$ for a specific connectivity function related to wireless communication networks, which is exponentially decaying with distance.

\begin{figure}[t]
\begin{center}
\includegraphics[scale=0.35]{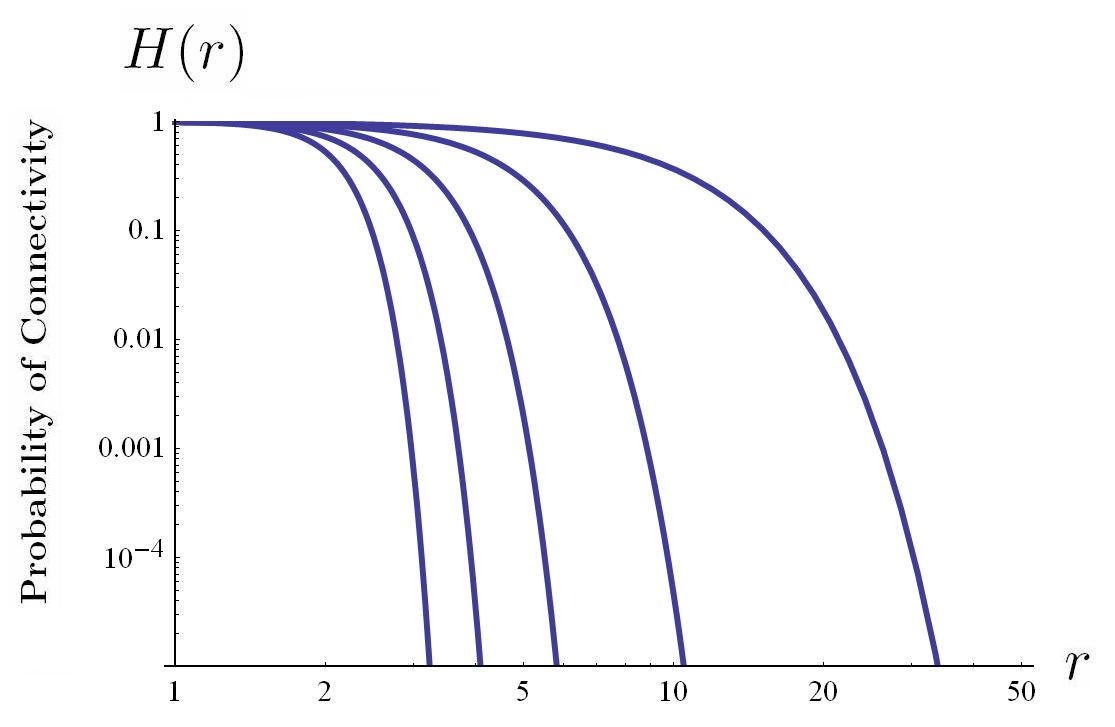}
\caption{\label{fig:H(r)} \footnotesize The probability of connecting a distance $r$ from a node is plotted for $\eta=2,3,4,5,6,$ form right to left, using $\beta=0.01$.}
\end{center}
\end{figure}

The information outage probability $P_{out}$ for a single-input single-output (SISO) Rayleigh fading link model\footnote{Other link models can also be considered. These may include single-input multiple-output (SIMO), multiple-input single-output (MISO), and multiple-input multiple-output (MIMO) (see also \cite{CDG11}).}, is given by \cite{tse2005fundamentals}
\es{P_{out}&=\mathrm{Pr} \pr{\log_{2}\pr{1+\mathrm{SNR} \times |h|^{2}} < x } \\
&=\mathrm{Pr}\pr {|h|^{2} < \frac{2^{x}-1 }{\mathrm{SNR} } },}
where $h$ is the channel transfer coefficient, $x$ is the minimum outage rate threshold, and the \textit{signal-to-noise ratio} (SNR) $\propto r^{-\eta}$ where $r$ is the dimensionless distance between connected nodes (relative to the signal wavelength), and $\eta$ is an environment dependent decay parameter\footnote{Typically $\eta = 2$ corresponds to propagation in free space but for practical reasons it is often modeled
as $\eta > 2$ for cluttered environments.}. The random variable $|h|^{2}$ is typically drawn from a \textit{standard exponential} distribution. Therefore the connectivity probability function can be written as
\es{\label{H}H(\mathbf{r})= 1- P_{out}= e^{-\beta r^{\eta}}
,}
with $\beta$ typically a small dimensionless constant, and is plotted in Figure~\ref{fig:H(r)} for $\beta=0.01$ and different values of $\eta$.
Notice that as $\eta\rightarrow\infty$ the connectivity is no longer probabilistic and converges to the interesting case of the popular unit disk model where connections have a fixed range of $r_{0}=\beta^{-\frac{1}{\eta}}$.

The surface element on a sphere of radius $R$ is
$R^2\sin\theta \mathrm{d}\theta d\phi$ in the usual spherical coordinates.  First we use the
Euclidean distance in $\R^{3}$ between two nodes, given in terms of the angle they subtend
at the center:
$d(\mathbf{r}_{1},\mathbf{r}_{2})= |\mathbf{r}_{12}|= \sqrt{2}R\sqrt{1-\cos\theta_{12}}$. Setting
$\eta=2$ (corresponding to propagation in free space) and locating one node at the pole $\theta=0$
we therefore have that
\es{\label{14} \int_{S^{2}} H(\mathbf{r}) \mathrm{d} \mathbf{r}&= \int_{0}^{2\pi}\int_{0}^{\pi} e^{-2 \beta R^{2} (1-\cos\theta)} R^{2}\sin\theta\mathrm{d}\theta \mathrm{d}\phi\\
&=\frac{\pi}{\beta}\prr{1-e^{-4\beta R^{2}}}\approx \frac{\pi}{\beta}
,}
when $\beta R^2$ (proportional to $V/V_H$ above) is large.

If instead geodesic (great-circle) distance is considered rather than Euclidean we have $d(\mathbf{r}_{1},\mathbf{r}_{2})= R\theta$ and hence
\es{\label{14b} \int_{S^{2}} H(\mathbf{r}) \mathrm{d} \mathbf{r}&= \int_{0}^{2\pi}\int_{0}^{\pi} e^{-\beta (R\theta)^{2}} R^{2}\sin\theta\mathrm{d}\theta \mathrm{d}\phi \\
&=\frac{\pi}{\beta}\prr{1- \frac{1}{\beta R^{2}}\pr{ \frac{1}{6}+ \frac{e^{-\beta \pi^{2}R^{2}}}{2\pi^{2}}} + \mathcal{O}\pr{\frac{1}{\beta^{2}R^{4}}} }  \approx\frac{\pi}{\beta}
,}
again when $\beta R^2$ is large. In this limit we see that the curvature ($R$-dependent)
effects become small for both distance functions.
It is thus possible to approximate the neighborhood of each node by a flat Euclidean metric,
more generally for other smooth manifolds $\mathcal{M}^d$, assuming a sufficiently fast decay
of $H(r)$.

Since the dominant contribution of $\int_{\mathcal{V}} H(\mathbf{r}_{1})H(\mathbf{r}_{12})\mathrm{d} \mathbf{r}_{1} $ is when nodes $1$ and $2$ are close to each other (\textit{i.e.} when $d(\mathbf{r}_{1},\mathbf{r}_{2})\ll R$), we adopt the above approximation in order calculate $K(\mathbf{r}_{2})$ using polar coordinates
\es{\label{15} K(\mathbf{r}_{2})&=\int_{S^{2}} H(\mathbf{r}_{1}) + H(\mathbf{r}_{12}) - H(\mathbf{r}_{1})H(\mathbf{r}_{12}) \mathrm{ d}\mathbf{r}_{1}\\
&\approx 2\frac{\pi}{\beta} - \int_{\R^{2}} H(\mathbf{r}_{1}) H(\mathbf{r}_{12})\mathrm{ d}\mathbf{r}_{1}  \\
&= 2\frac{\pi}{\beta} - \int_{0}^{\infty}\int_{0}^{2\pi} r_{1} e^{-\beta (2r_{1}^{2} + r_{2}^{2} - 2r_{1}r_{2} \cos\theta)} \mathrm{ d}\theta \mathrm{ d}r_{1}  \\
&= 2\frac{\pi}{\beta} - \int_{0}^{\infty} 2\pi r_{1} e^{-\beta (2r_{1}^{2} + r_{2}^{2})} I_{0}(2\beta r_{1}r_{2}) \mathrm{ d}r_{1}  \\
&= 2\frac{\pi}{\beta} - \frac{\pi}{2\beta}e^{-\beta\frac{r_{2}^{2}}{2}}
,}
where $I_{0}(x)$ is the modified Bessel function of the first kind, $r_{1}$ and $r_{2}$ are the
distances of the two nodes from the origin and $\theta$ the angle between them.
Finally, we may evaluate
\es{\label{16} \int_{S^{2}} H(\mathbf{r}) e^{-\rho K(\mathbf{r})}  \mathrm{ d}\mathbf{r} &\approx \int_{\R^{2}} H(\mathbf{r}) e^{-\rho K(\mathbf{r})}  \mathrm{ d}\mathbf{r}\\
&= e^{-\frac{3\pi \rho}{2\beta}}\pr{\frac{2}{\pi\rho} + \mathcal{O}(\rho^{-2})}
,}
by expanding the exponential in equation~\eqref{15} around $r_{2}=0$ as this is near the main contribution to the integral.
Altogether, according to equation~\eqref{13} this gives
\es{\label{2nd} P_{fc} & \approx  1- N e^{-\rho \frac{\pi}{\beta}} -  \frac{N}{\pi} e^{-\frac{3\pi \rho}{2\beta}}
,}
to leading order.

Note that the above calculation can of course be performed for general $\eta>0$ and $n$-sphere.  The
general case is excluded for the sake of brevity, however
one case that provides some insight is the ``unit-disk''
model, corresponding to the limit $\eta\to\infty$ with
$\beta$ scaled so that $\beta^{-1/\eta}$ approaches a
limit, $r_0$.  We then have
\es{
H(r)=\left\{\begin{array}{cc}1&r\leq r_0\\0&r>r_0\end{array}\right.}
leading to
\es{
\int_{\mathbb{S}^2}H({\mathbf r})\mathrm{d}{\mathbf r}
\approx \int_{\mathbb{R}^2}H({\mathbf r})\mathrm{d}{\mathbf r}=2 \pi r_{0}^{2}}
for either metric (actually it is exact for the Euclidean
metric).  We then calculate on $\mathbb{R}^2$ for
$r_2<2r_0$
\es{
K({\bf r}_2)&=2\pi r_0^2+\frac{r_2}{2}\sqrt{4r_0^2-r_2^2}
-2r_0^2\arccos\frac{r_2}{2r_0}\\
&=\pi r_0^2+2r_0r_2-\frac{r_2^3}{12r_0}+ \ldots}
expanding for small $r_2$ as this leads to the main
contribution to $P_{fc}$.  The latter comes to
\es{
P_{fc}= 1-Ne^{-\pi r_0^2\rho}\left[
1+\frac{\pi}{4r_0^2\rho}+\ldots\right]}
where the last term is the second order contribution.

We now make an important observation. From equation~\eqref{2nd} we notice that in the high density limit the second order term in our expansion of $P_{fc}$ has exponent of order $\sim -\frac{3\pi \rho}{2\beta}$ and hence is exponentially smaller than the first order term. This is due to the probabilistic nature
of the connections, two nodes need more excluded
volume on average to be isolated than a single node.
For the unit disk model, however, the two nodes can
be very close with a probability that is an algebraic function of their mutual distance, requiring roughly the same excluded volume, and leading to a second
order effect which differs from the first only by an algebraic function.

This means that higher order effects appear to be
important mostly for the unit disk model. For the
probabilistic model which is more realistic for
applications such as wireless networks, we are justified to use only the first order results for the discussion
of boundary effects in confined geometries; to this
we now turn.

\section{Boundary effects}

In this section, we lift the homogeneity assumption and consider the first and second order approximations of $P_{fc}$ only to discover that contrary to popular belief and practice, boundary
effects matter. This is a central observation of the current work, which we detail and discuss in
the following subsections.

\subsection{Inhomogeneous first and second order approximations}

Returning to the first order equation~\eqref{4} but no longer assuming homogeneity gives
\es{\label{7} P_{fc} & \approx 1-\frac{N}{V}\int_{\mathcal{V}} \pr{1-\frac{1}{V}\int_{\mathcal{V}} H(\mathbf{r}_{12}) \mathrm{ d}\mathbf{r}_{1} }^{N-1} \mathrm{ d}\mathbf{r}_{2}\\
&= 1-\rho \int_{\mathcal{V}} e^{-\rho\int_{\mathcal{V}} H(\mathbf{r}_{12}) \mathrm{ d}\mathbf{r}_{1} }\pr{1+ \mathcal{O}(N^{-1})} \mathrm{ d}\mathbf{r}_{2}
,}
for large $N$. This equation was recently given in \cite{Mao11} (equation (8)) with $V$ scaled exponentially with $\rho$ thus ignoring any boundary effects. Recall that for this approximation
we only require that $V\gg\rho$ or equivalently $V\gg \sqrt{N}$. This is a key difference from previous percolation approaches.

Similarly, from equations~\eqref{9} and~\eqref{11b}, one can show that the second order approximation of $P_{fc}$ becomes
\es{\label{7b} P_{fc} &= 1-\rho \int_{\mathcal{V}} e^{-\rho\int_{\mathcal{V}} H(\mathbf{r}_{12}) \mathrm{ d}\mathbf{r}_{1} }\mathrm{ d}\mathbf{r}_{2}
- \frac{N(N-1)}{2V^{2}}\int\int_{\mathcal{V}} H(\mathbf{r}_{12}) \pr{1- \frac{\hat{K}(\mathbf{r}_{1},\mathbf{r}_{2})}{V} }^{N-2} \mathrm{ d}\mathbf{r}_{1} \mathrm{ d}\mathbf{r}_{2}\\
&\approx 1-\rho \int_{\mathcal{V}} e^{-\rho\int_{\mathcal{V}} H(\mathbf{r}_{12}) \mathrm{ d}\mathbf{r}_{1} }\mathrm{ d}\mathbf{r}_{2}
- \frac{\rho^{2}}{2} \int\int_{\mathcal{V}} H(\mathbf{r}_{12}) e^{-\rho \hat{K}(\mathbf{r}_{1},\mathbf{r}_{2}) } \mathrm{ d}\mathbf{r}_{1} \mathrm{ d}\mathbf{r}_{2}
,}
where
\es{\hat{K}(\mathbf{r}_{1},\mathbf{r}_{2})=  V - \int_{\mathcal{V}}(1-H(\mathbf{r}_{13}))(1-H(\mathbf{r}_{23})) \mathrm{ d}\mathbf{r}_{3}
,}
and in the last step of \eqref{7b} we have ignored terms of order $\sim N^{-1}$ since $N$ is also assumed to be large.

Equation~\eqref{7} suggests that in the high density limit, the probability of having a single $N-1$ connected cluster is dominated by nodes which are situated in ``\textit{hard to connect}" regions of the available domain $\mathcal{V}$. This is because the outer integral in \eqref{7} is dominated by
contributions where the integral in the exponential is small, for example at corners, edges and faces. The second order corrections in~\eqref{7b}, as expected from our calculations in section 2.4, are of secondary importance and do not offer further insight.
It is important to note that our approach here contradicts the usual universality scenario found in statistical mechanics and emphasizes the dominant importance of boundaries (and in particular corners). We stress here that this observation does not depend on using Euclidean distance and is valid in any geometry and any dimension where the lack of connectivity is due to a situation involving a single disconnected node and an $N-1$ cluster.. In such a case, the outage probability will be dominated by situations where that node has a small volume in range.

Returning to the general Rayleigh fading link model of equation~\eqref{H} we now demonstrate the above observation in a straight forward way. Suppose that $\mathbf{r}_{2}$ is situated somewhere on the boundary of the network domain $\mathcal{V}\subset \R^{d}$. Since direct connectivity is exponentially decaying, we may separate the $\mathrm{d} \mathbf{r}_{1}$ integral and extract its leading order behavior
\es{\label{sep} \int_{\mathcal{V}}H(\mathbf{r}_{12})\mathrm{d}\mathbf{r}_{1}&\approx\pr{\int_{0}^{\infty}r^{d-1}e^{-\beta r^{\eta}}\mathrm{d}r}\pr{\int\mathrm{d}\Omega}= \frac{\Gamma\pr{\frac{d}{\eta}}}{\eta \beta^{\frac{d}{\eta}}}\omega,
}
where $\Omega=\frac{2\pi^{\frac{d}{2}}}{\Gamma\pr{\frac{d}{2}}}$ is the full solid angle in $d$ dimensions and $\omega\in(0, \Omega)$ is the solid angle available from $\mathbf{r}_{2}$. In the case of $\eta=2$ the gamma functions cancel and equation~\eqref{sep} simplifies to $\pr{\frac{\pi}{\beta}}^{\frac{d}{2}}\frac{\omega}{\Omega}$. As disused before, scaling $\beta=r_{0}^{-\eta}$ we recover the popular unit-disk model in the limit of $\eta\rightarrow\infty$, allowing for direct comparison with the probabilistic case of \eqref{H} and other earlier results. For example, if $\eta=d$ then equation \eqref{sep} gives $r_{0}^{d}/d$ which is also what the unit-disk model gives.

In general we expect that the outer integral in \eqref{7} will have contributions from boundary regions as in \eqref{sep}, with a term of the order of
$\exp(-\rho\omega\int H(r) r^{d-1}dr)$ for the smallest $\omega$, for example the pointiest corner,
dominating $P_{fc}$ at high density. It is clear from this argument that connectivity is indeed dominated from regions which are hard to connect and is controlled by the solid angle available to them.

The above observation has brought forward a radically different understanding of connectivity in confined geometries, namely that full connectivity is dominated by the critical, hard to connect areas such as corners, edges and faces. Furthermore, this suggests the decomposition of the probability of a fully connected network $P_{fc}$ into a sum of contributions due to boundary objects with different solid angles.
In order to obtain a more in-depth understanding of this novel idea we now embark into a more detailed investigation by considering a few example domains.
Through these examples, the boundary effects will become clear thus leading to the main result of this section; a general formula (see equation~\eqref{gen1}) for $P_{fc}$.

\subsection{Example 1. Circle ($\eta=2$)}

We start with a circular domain $c_{R}\subset\R^{2}$ of radius $R$. Using a Euclidean metric, the distance between two nodes is given by $d(\mathbf{r}_{1},\mathbf{r}_{2})= |\mathbf{r}_{12}|= \sqrt{|\mathbf{r}_{1}|^{2}+|\mathbf{r}_{2}|^{2}-2 |\mathbf{r}_{1}| |\mathbf{r}_{2}|\cos\theta}$. Hence, for the case of $\eta=2$ we must first evaluate
\es{\label{23} \int_{c_{R}} H(\mathbf{r}_{12}) \mathrm{ d}\mathbf{r}_{1}&= \int_{0}^{R}\int_{0}^{2\pi} \pr{r_{1} e^{-\beta\pr{r_{1}^{2}+r_{2}^{2}-2r_{1} r_{2}\cos\theta}}} \mathrm{ d}\theta \mathrm{ d}r_{1}\\
&= 2\pi \int_{0}^{R} \pr{r_{1} I_{0}( 2 r_{1} r_{2} \beta) e^{-\beta(r_{1}^{2}+r_{2}^{2})} }\mathrm{ d}r_{1},}
where $I_{0}(x)$ is the modified Bessel function of the first kind. Unlike the ordinary Bessel function, which is oscillatory for real arguments, $I_{0}(x)$ is exponentially growing.
When $\mathbf{r}_{2}$ is close to the center (\textit{i.e.} $r_{2} \approx 0$) we have that $e^{-\beta r_{2}}I_{0}(2 r_{1} r_{2} \beta)= 1 + \mathcal{O}(r_{2}^{2})$ so that the integral becomes:
\es{2\pi \int_{0}^{R} \pr{r_{1} I_{0}( 2 r_{1} r_{2} \beta) e^{-\beta(r_{1}^{2}+r_{2}^{2})} }\mathrm{ d}r_{1}= \frac{\pi}{\beta}\pr{1-e^{-\beta R^{2}}} + \mathcal{O}(r_{2}^{2}) \approx \frac{\pi}{\beta},}
where we have assumed that $\beta R^{2}\gg1$. Notice that this is the same assumption made in equations~\eqref{14} and~\eqref{14b} and will be consistently used throughout our calculations.
Otherwise, if $\mathbf{r}_{2}$ is not near the center of $c_{R}$, we have the asymptotic relation $I_{0}(x)= \frac{e^{x}}{\sqrt{2\pi x}}\pr{1+\mathcal{O}(x^{-1})}$ so that the integral becomes:
\es{ 2\pi \int_{0}^{R} \pr{r_{1} I_{0}(2 r_{1} r_{2} \beta) e^{-\beta(r_{1}^{2}+r_{2}^{2})} }\mathrm{ d}r_{1} &= 2\pi \int_{0}^{R} \pr{r_{1} \frac{e^{2 r_{1} r_{2} \beta}}{\sqrt{4\pi r_{1} r_{2} \beta}} e^{-\beta(r_{1}^{2}+r_{2}^{2})} }\mathrm{ d}r_{1}  \\
&\approx \frac{\sqrt{\pi}}{\sqrt{\beta}} \int_{0}^{R} e^{-\beta(r_{1}-r_{2})^{2}} \mathrm{ d}r_{1}\\
&= \frac{\pi}{2\beta}\pr{\mathrm{erf}\prr{\sqrt{\beta}(R-r_{2})}  + \mathrm{erf}\prr{\sqrt{\beta}r_{2}}}
,}
since the main contribution of this integral comes from $r_{1}\approx r_{2}$.
Matching the two solutions for small and large $r_{2}$ we obtain a nice approximation to equation~\eqref{23}
\es{ \int_{c_{R}} H(\mathbf{r}_{12}) \mathrm{ d}\mathbf{r}_{1}&\approx\frac{\pi}{2\beta} \pr{\mathrm{erf}\prr{\sqrt{\beta}(R-r_{2})}  + 1}\\
&= \frac{\pi}{2\beta}f(r_{2}).}

In order to make progress, we approximate $f(r)$ by
\es{\label{approx} \tilde{f}(r)=\begin{cases}
c_{1}, &   \text{for} \quad 0 < r < a,\\
c_{2} - m (r-R), &   \text{for} \quad a \leq r < R, \end{cases}
}
where $c_{1}= 2\mathrm{erf}\prr{\sqrt{\beta}\frac{R}{2}}\approx 2$, $c_{2}= \mathrm{erf}\prr{\sqrt{\beta}R }\approx1$, $m= \frac{2 \sqrt{\beta}}{\sqrt{\pi}}\pr{1-e^{-\beta R^{2}}} \approx \frac{2 \sqrt{\beta}}{\sqrt{\pi}}$ when $\sqrt{\beta}R\gg1$ and $a$ is their intersection approximating the location of the effective turning point of $f(r)$.

The outer integral in equation~\eqref{7} can now be approximated to give
\es{ \label{27} P_{fc}&=1-\rho\int_{c_{R}} e^{-\rho\int_{c_{R}} H(\mathbf{r}_{12}) \mathrm{ d}\mathbf{r}_{1} } \mathrm{ d}\mathbf{r}_{2} \\
&\approx 1-2\pi \rho \int_{0}^{R} r e^{-\rho \frac{\pi}{2\beta}\tilde{f}(r) } \mathrm{ d}r \\
&= 1- \pi R^{2} \rho e^{-\rho \frac{\pi}{\beta}} \pr{1-\frac{\sqrt{\pi}}{\sqrt{\beta}R} - \frac{2\sqrt{\beta}}{\rho\sqrt{\pi}R} + \mathcal{O}(R^{-2})} - 2\pi R \sqrt{\frac{\beta}{\pi}} e^{-\frac{\pi}{2 \beta}\rho}\pr{1-\frac{\sqrt{\beta}}{\rho\sqrt{\pi}R} + \mathcal{O}(R^{-2})}
.}

\begin{figure}[t]
\begin{center}
\includegraphics[scale=0.32]{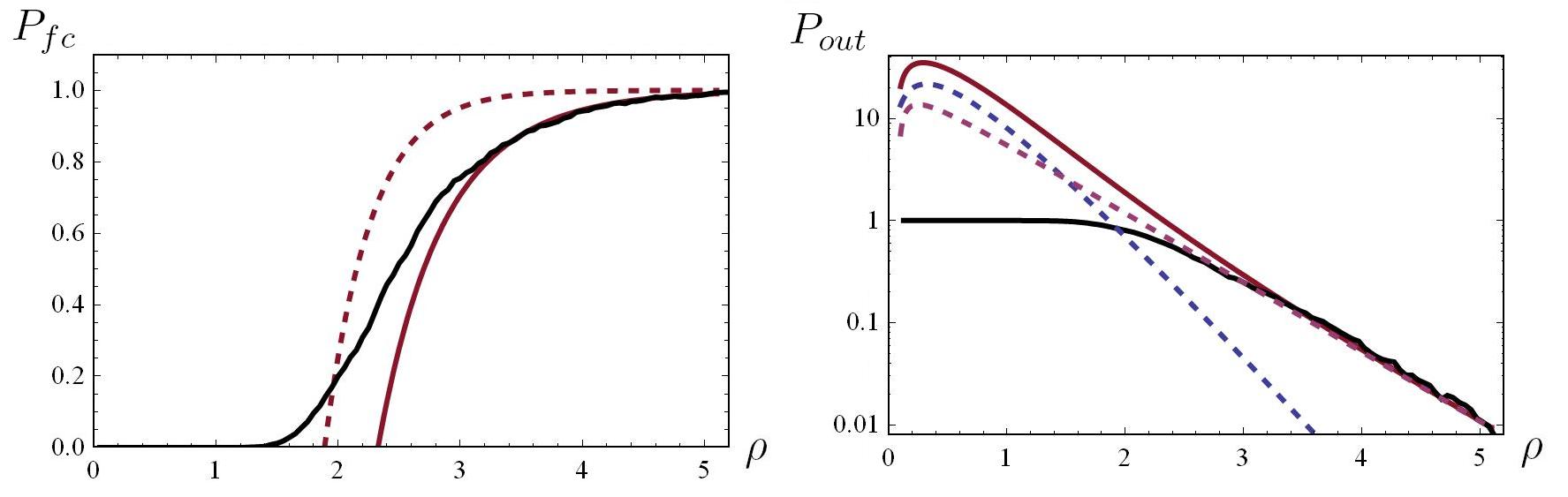}
\caption{\label{fig:circular} \footnotesize \emph{Left:} (Color online) A plot of $P_{fc}(\rho)$ in the circular domain $c_{R}$ of area $A=200$ ($\beta=1$) (solid red) compared with the prediction of equation~\eqref{6} (dashed) and a numerical simulation (black). \emph{Right:} The dashed curves show the contributions from the bulk (blue) and boundary (purple) term in equation~\eqref{27} on a log-linear scale. The full curves are plots of $P_{out}(\rho)$ with dark red and black corresponding to theory and numerical simulation.}
\end{center}
\end{figure}

It is clear from equation~\eqref{27} that $P_{fc}$ is composed of a bulk term and a boundary term with coefficients characteristic of the area and perimeter of $c_{R}$, with corrections due to the curvature of the boundary which vanish as $R\rightarrow\infty$. Moreover, $P_{fc}$ is dominated by the boundary term in the high density limit. This can be seen on the right panel of Figure~\ref{fig:circular}, and that this transition occurs at $\rho_{t} \sim {\frac{2\beta}{\pi}} \ln \frac{\sqrt{\pi}}{2\sqrt{\beta}}R $ as $R\rightarrow\infty$.

The left panel of Figure~\ref{fig:circular} shows an encouraging comparison with a numerical Monte Carlo simulation. Nodes are randomly placed within a circular disc of area $A=200$ and are connected with probability $H(\mathbf{r}_{ij})$ ($\beta=1$). The error at lower densities is of course expected as the theory developed above is only a first order approximation to $P_{fc}$. Nevertheless, the agreement with simulations substantially improves at high densities but is also good in mid-densities. Hence, equation~\eqref{7} provides a useful path which one may take to predict, control, or even optimize and set benchmarks for achieving full network connectivity.

\subsection{Example 2. Square ($\eta=2$)}

We now consider a square domain $s_{L}$ of side $L$. The integral in the exponent of~\eqref{7} can be evaluated exactly in Cartesian coordinates for $\eta=2$ to give
\es{\int_{s_{L}} H(\mathbf{r}_{12}) \mathrm{ d}\mathbf{r}_{1}&=  \int_{-\frac{L}{2}}^{\frac{L}{2}} \int_{-\frac{L}{2}}^{\frac{L}{2}}  \pr{e^{-\beta \pr{(x_{1}-x_{2})^{2}+(y_{1}-y_{2})^{2} }}} \mathrm{ d}x_{1} \mathrm{ d}y_{1}\\
&= \frac{\pi}{4\beta} h(x_{2})h(y_{2})
,}
where $h(x)= \pr{\mathrm{erf}\prr{\sqrt{\beta}\frac{(L-2 x)}{2}} + \mathrm{erf}\prr{\sqrt{\beta}\frac{(L+2 x)}{2}} }$.
Due to symmetry, we need only consider the positive quadrant of $s_{L}$ \textit{i.e.} when $x_{2},y_{2}\in [0,\frac{L}{2}]$. Hence $h(x)\approx \pr{\mathrm{erf}\prr{\sqrt{\beta}\frac{(L-2 x)}{2}}+ 1}$ and may be further approximated by $\tilde{f}(x)$ with $R$ replaced by $L/2$ (see equation~\eqref{approx}) such that
\es{\label{30}\int_{0}^{\frac{L}{2}}\int_{0}^{\frac{L}{2}} e^{-\rho  \frac{\pi}{4\beta} h(x)h(y)} \mathrm{ d}x \mathrm{ d}y &\approx
\int_{0}^{\frac{L}{2}}\int_{0}^{\frac{L}{2}} e^{-\rho  \frac{\pi}{4\beta} \tilde{f}(x)\tilde{f}(y)} \mathrm{ d}x \mathrm{ d}y
.}
Note however that it is crucial that the quadratic $x_{2} y_{2}$ term is not included in the expansion of $h(x_{2})h(y_{2})$ as it is of higher order. Performing the integrals and multiplying by four we obtain a result
\es{\label{31} P_{fc} &=1- L^{2}\rho e^{-\frac{\pi}{\beta} \rho }\pr{1-\mathcal{O}(L^{-1})} - 4 L\sqrt{\frac{\beta}{\pi}}  e^{-\frac{\pi}{2\beta} \rho}\pr{1-\mathcal{O}(L^{-1})} - \frac{16\beta}{\rho\pi}  e^{-\frac{\pi}{4\beta} \rho}
,}
assuming that $\sqrt{\beta}\frac{L}{2}\gg1$. Note that the above calculation can naturally be extended in a straight forward way to include for general $\eta>0$ and higher dimensional $d>0$ orthotopes (hyper-rectangles) of non-equal sides.

\begin{figure}[t]
\begin{center}
\includegraphics[scale=0.32]{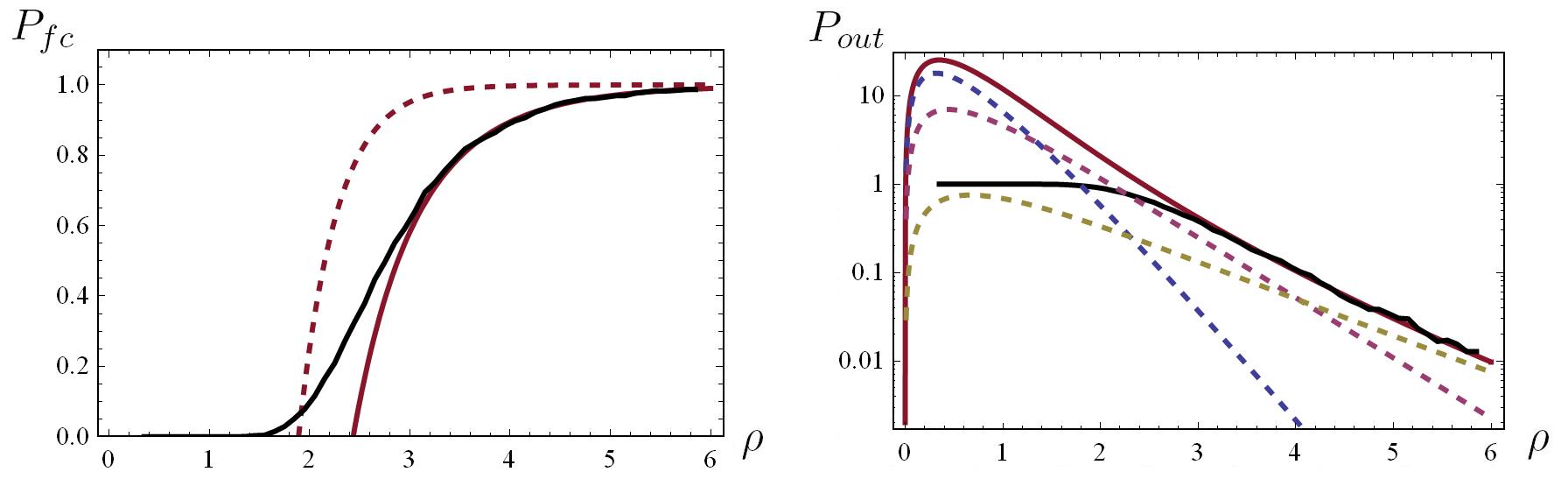}
\caption{\label{fig:square} \footnotesize (Color online) \emph{Left:} A plot of $P_{fc}(\rho)$ in the square domain $s_{L}$ of area $A=200$ ($\beta=1$) (solid red) compared with the prediction of equation~\eqref{6} (dashed) and a numerical simulation (black). \emph{Right:} The dashed curves show the contributions from the bulk (blue), the boundary (purple) and the corner (yellow) term in equation~\eqref{31} on a log-linear scale. The full curves are plots of $P_{out}(\rho)$ with dark red and black corresponding to theory and numerical simulation.}
\end{center}
\end{figure}

As expected, $P_{fc}$ in equation~\eqref{31} is composed of a bulk term, a boundary and a corner term with coefficients characteristic of the corresponding volume of the relevant object (\textit{e.g.} area, perimeter, number of corners) and is dominated by the corner term in the high density limit. This is illustrated on the right panel of Figure~\ref{fig:square}. One would expect however the coefficient of the corner term to be $4$ rather than $16$. To understand why this is not so, we will consider a general angle term in the next subsection. Finally, we notice that there exists a parameter window $\rho_{t_{1}} \leq \rho \leq \rho_{t_{2}}$ where the edges of the square provide the dominant contribution to $P_{fc}$. The left panel of Figure~\ref{fig:square} shows another encouraging comparison with numerical simulations.

\subsection{Example 3. Circle sector: General angle ($\eta>0$)}

In order to understand corner effects better, we now consider for general $\eta>0$ a circular sector $c_{R,\vartheta}\subset \R^{2}$ of central angle $\vartheta\in(0,\pi)$ and circle radius $R$. We ignore the bulk and boundary contributions to $P_{fc}$, and concentrate on the corner term which we denote $\mathcal{C}_{d}(\vartheta)$ for $d=2$. Note that the distance between two nodes is now given by $d(\mathbf{r}_{1},\mathbf{r}_{2})= \sqrt{|\mathbf{r}_{1}|^{2}+|\mathbf{r}_{2}|^{2}-2 |\mathbf{r}_{1}| |\mathbf{r}_{2}|\cos(\theta_{1}-\theta_{2})}$ with angular positions $\theta_{1,2}\in(0,\vartheta)$. Following the arguments of section 3.1, we expect that to leading order $\mathcal{C}_{2}(\vartheta)$ is when $\mathbf{r}_{2}$ is situated near the corner \textit{i.e.} when $r_{2}\approx 0$.
Hence we expand $H(\mathbf{r}_{12})$ about the corner in polar coordinates such that to leading order
\es{\label{38} \int_{c_{R,\vartheta}} H(\mathbf{r}_{12}) \mathrm{d} \mathbf{r}_{1}&= \lim_{R\rightarrow\infty}\int_{0}^{\vartheta}\int_{0}^{R} r_{1} e^{-\beta r_{1}^{\eta}}\pr{1+\eta \beta r_{1}^{\eta-1} r_{2} \cos(\theta_{1}-\theta_{2}) + \mathcal{O}(r_{2}^{2})}   \mathrm{d} r_{1} \mathrm{d} \theta_{1}\\
&= \frac{\vartheta \Gamma\pr{\frac{2}{\eta}} + r_{2} \Gamma\pr{\frac{1}{\eta}}\pr{\sin\theta_{2} -\sin(\theta_{2}-\vartheta) } }{\eta \beta^{\frac{2}{\eta}}}+ \mathcal{O}(r_{2}^{2})
,}
and hence
\es{\label{C2} \mathcal{C}_{2}(\vartheta)&= \lim_{R\rightarrow\infty} \int_{0}^{\vartheta} \int_{0}^{R} r_{2} e^{-\rho\pr{\frac{\vartheta \Gamma\pr{\frac{2}{\eta}} + r_{2} \Gamma\pr{\frac{1}{\eta}}\pr{\sin\theta_{2} -\sin(\theta_{2}-\vartheta) } }{\eta \beta^{\frac{2}{\eta}}}+ \mathcal{O}(r_{2}^{2}) }}  \mathrm{d} r_{2} \mathrm{d}\theta_{2}\\
&= \frac{\beta^{\frac{2}{\eta}} e^{-\rho \frac{\Gamma\pr{\frac{2}{\eta}}}{\eta \beta^{\frac{2}{\eta}}} \vartheta} }{\rho^{2} \sin\vartheta  \Gamma\pr{1+\eta^{-1}}^{2}}
.}

\begin{figure}[t]
\begin{center}
\includegraphics[scale=0.35]{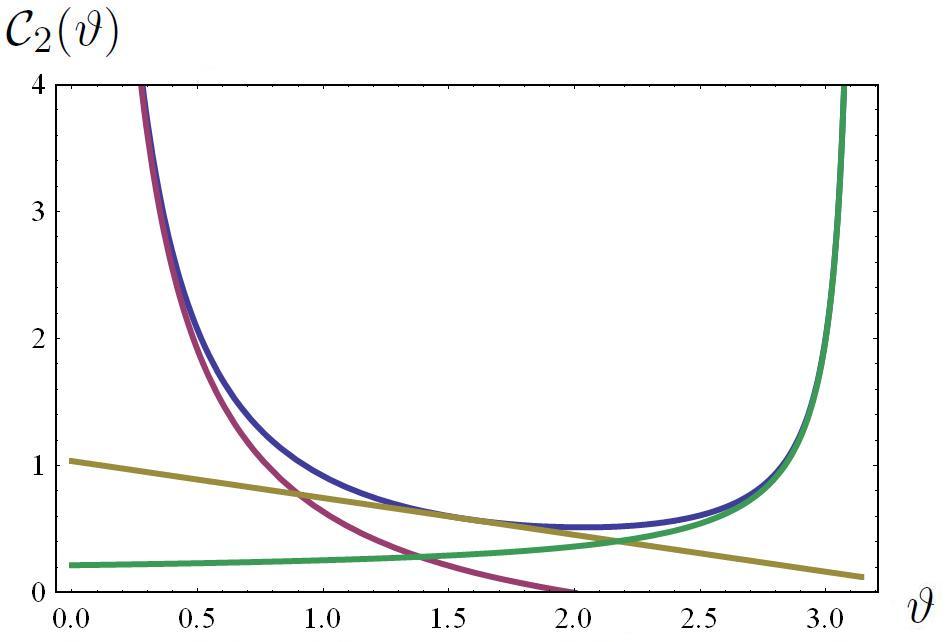}
\caption{\label{fig:C(m)} \footnotesize (Color online) The contribution from a corner of angle $\vartheta$ (blue) for $\rho=\beta=1$ and $\eta=2$. Also shown are the three asymptotic regimes (see equation~\eqref{c2}) of $\vartheta\rightarrow 0,\frac{\pi}{2},\pi$ shown in purple, yellow and green respectively.}
\end{center}
\end{figure}

Reassuringly in the case of a square with $\eta=2$ we have that $4 \mathcal{C}_{2}(\frac{\pi}{2})= \frac{16\beta}{\rho^{2}\pi}  e^{-\frac{\pi}{4\beta} \rho}$ as in equation~\eqref{31}. Furthermore, from \eqref{C2} we identify three main regimes as follows
\es{\label{c2} \mathcal{C}_{2}(\vartheta) = \begin{cases}
\frac{4\beta}{\pi\rho^{2} \vartheta } -\frac{2}{\pi\rho}+ \mathcal{O}(\vartheta), &   \text{for}\quad  \vartheta \ll 1,\\
\frac{4\beta}{\pi\rho^{2}} e^{\frac{-\pi\rho}{4\beta}} - \frac{2}{\pi\rho}e^{\frac{-\pi\rho}{4\beta}}\pr{\vartheta-\frac{\pi}{2}}+\mathcal{O}((\vartheta-\frac{\pi}{2})^2), &   \text{for} \quad \vartheta \approx \frac{\pi}{2},\\
\frac{4\beta e^{\frac{-\pi \rho}{2\beta}}}{\pi\rho^{2} (\pi-\vartheta)} +\frac{2e^{\frac{-\pi\rho}{2\beta}}}{\pi\rho}+\mathcal{O}(\pi-\vartheta), &   \text{for}\quad  \pi-\vartheta \ll 1,\\
 \end{cases}
}
for the case of $\eta=2$ and are illustrated in Figure~\ref{fig:C(m)}. Finally, $\mathcal{C}_{2}(\vartheta)$ has a minimum at
\es{\vartheta_{min}=\pi-\arctan\frac{2\beta}{\rho}}

Physically, the picture shown in Figure~\ref{fig:C(m)} makes sense in the following way: as $\vartheta\rightarrow\pi$, the corner turns into a wall of infinite length and thus $\mathcal{C}_{2}(\vartheta)$ diverges. This argument is supported by the fact that in this limit, the exponent is $\sim -\frac{\pi\rho}{2\beta}$ and hence of the same order as that of the perimeter term in equation~\eqref{31}.

\subsection{A General formula}
\begin{figure}[t]
\begin{center}
\includegraphics[scale=0.32]{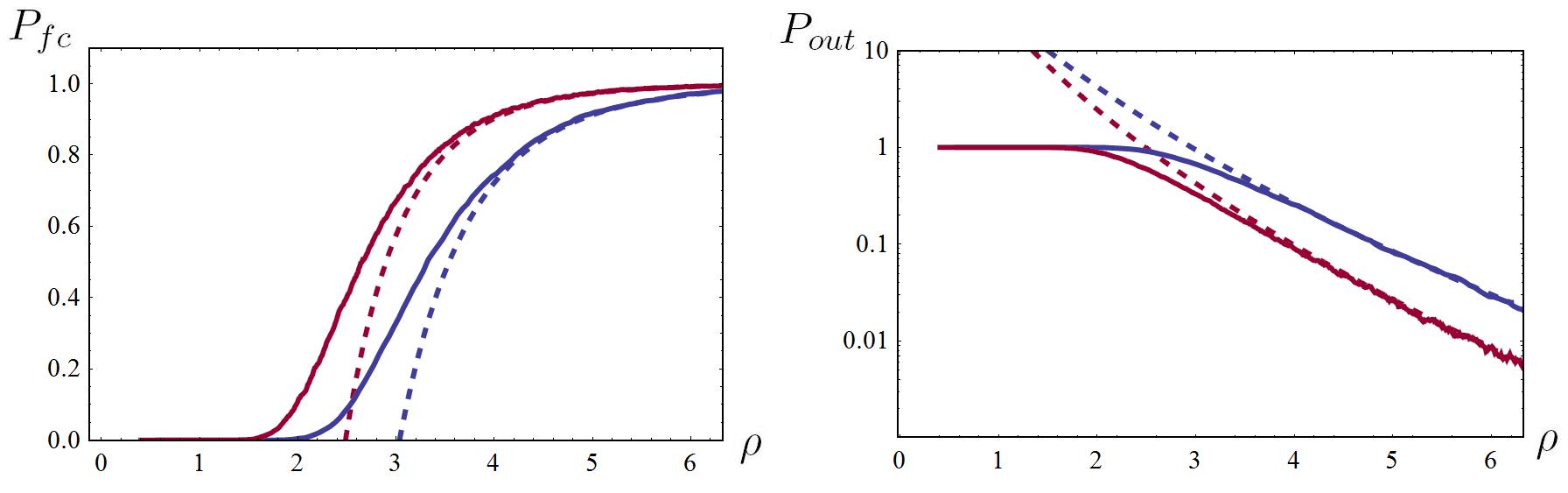}
\caption{\label{fig:cube} \footnotesize \emph{Left:} (Color online) Comparison of $P_{fc}(\rho)$ from theoretical predictions of \eqref{gen1} (dashed curves) with direct numerical simulation (jagged line) of two random graphs inside a cube of side $L=7$ for the unit-disk connectivity function with $r_{0}=1$ (blue) and the probabilistic Rayleigh fading link model with $\eta=2$ (red). \emph{Right:} The corresponding $P_{out}(\rho)$ is plotted on a log-linear scale emphasizing the good agreement between theory and simulations as well as the difference between the two models.}
\end{center}
\end{figure}

In high-dimensional domains $\mathcal{V}\subset\R^{d}$ ($d\geq3$), boundary effects are due to intersections of hyperplanes which the above examples and techniques can generalize to account for. As a result $P_{fc}$ can be expressed as a sum of contributions due to objects of different codimension $i=0,1,\ldots d$ (with $i=0$ corresponding to the bulk/volume term)
\es{\label{gen1}P_{fc}  \approx  1- \sum_{i=0}^{d}\sum_{j_{i}} \rho^{1-i} \mathbf{G}_{j_{i}} \mathbf{V}_{j_{i}} e^{-\rho \omega_{j_{i}} \mathcal{H} }
,}
where $d\in\N$ is the space dimension, $\mathbf{G}_{j_{i}}$ is a geometrical factor for each object $j$ of codimension $i$ while $\mathbf{V}_{j_{i}}$ is the corresponding $d-i$ dimensional volume of the object with solid angle $\omega_{j_{i}}\in(0,\Omega)$ and $\mathcal{H}=\int_{0}^{\infty}r^{d-1}H(r)\mathrm{d}r$. It is clear that for $i=0$ and $1$, the second sum in~\eqref{gen1} contains only one term. The remaining one dimensional integral is easily computed for typical connectivity functions $H_{ij}$ (also see \cite{CDG11}). Corrections due to curved hyper-surfaces are expected to give algebraic corrections as in \eqref{27}. Note however that $\mathbf{G}_{j_{i}}$ is $H_{ij}$ dependent while $\mathbf{V}_{j_{i}}$ is not. In the case of $H(r)=e^{-\beta r^{2}}$ we find for example for $d$-dimensional hyper-rectangles where all hyperplanes meet at right angles $\mathbf{G}_{j_{i}} = 2^{i(i-1)}\pr{\frac{\beta}{ \pi}}^{\frac{i(d-1)}{2}}$ while for the unit-disk model with range $r_{0}$ we find $\mathbf{G}_{j_{i}} = 2^{i(i-1)}/V_{d-1}^{i}$, where $V_{n}=\frac{\pi^{n/2}r_{0}^{n}}{\Gamma(n/2+1)}$ is the volume of an $n$-dimensional sphere of radius $r_{0}$. These two cases are compared in Figure~\ref{fig:cube} for a cube of side $L=7$.

\begin{figure}[t]
\begin{center}
\includegraphics[scale=0.32]{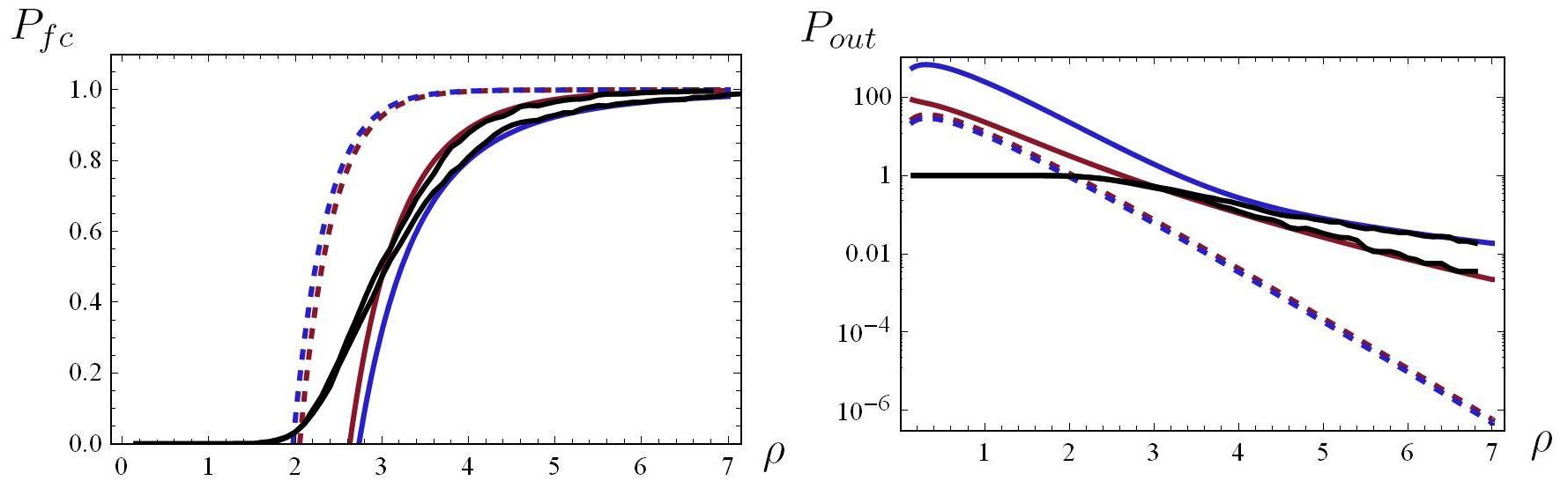}
\caption{\label{fig:sector} \footnotesize (Color online) \emph{Left:} Comparison of $P_{fc}(\rho)$ from theoretical predictions of \eqref{gen1} (solid curves) with direct numerical simulation (black) of two random graphs inside circular sectors $c_{R_{1},\vartheta_{1}}$ (blue) and $c_{R_{2},\vartheta_{2}}$ (red) using $\beta=1$ and $\eta=2$. The sectors are of equal perimeters $\mathbf{V}_{1_{1}}=70$ but with $\vartheta_{1}=\pi/4$ and $\vartheta_{2}=3\pi/4$. The dashed curves correspond to conventional wisdom (\textit{i.e.} the bulk contribution of equation~\eqref{6}). \emph{Right:} The corresponding $P_{out}(\rho)$ is plotted on a log-linear scale emphasizing the good agreement between theory and simulations.}
\end{center}
\end{figure}
The above formula is a high density expansion giving the probability of achieving a fully connected network of nodes in confined geometries. For all practical purposes ($d\leq3$) equation~\eqref{gen1} can account for most interesting (from an engineering point of view) domains in which a network may be deployed.
Furthermore, its closed and simple format emphasizes the logical decomposition of the domain into objects of different full connectivity importance. This is illustrated in Figure~\ref{fig:sector} comparing two circular sectors ($d=2$) of angles $\vartheta_{1}=\pi/4$ and $\vartheta_{2}=3\pi/4$ but of equal perimeters $\mathbf{V}_{1_{1}}=70$. Notice that the density at which the numerical simulation results become distinct ($\rho\approx3$) in the left panel of Figure~\ref{fig:sector} is also where the corner term of $c_{R_{1},\vartheta_{1}}$ overtakes the common perimeter term (shown in black) in the right panel of Figure~\ref{fig:sector}.

\section{Conclusion}

In this paper we have attempted to understand how boundary effects can influence full connectivity in confined geometries. Based on the fact that at very high densities the probability of full connectivity $P_{fc}$ is simply the complement of the probability of a single isolated node, we have developed a cluster expansion approach which classifies terms with respect to their largest connected component. Assuming a homogeneous system and averaging over the available space of the domain, we extract the leading and next-to-leading order behavior of $P_{fc}$ as is given in equations~\eqref{6} and~\eqref{13} respectively. Although our approach can be consistently expanded to obtain even higher order corrections to $P_{fc}$, using an exponentially decaying connectivity test-function\footnote{This was taken from a standard Rayleigh fading link model (also see \cite{CDG11}).} (equation~\eqref{H}), we showed that correction terms are exponentially small. This is because nodes are `connected in probability' (in this case exponentially) rather than in an all-or-nothing way.

The first main result of this paper follows from lifting the homogeneity assumption to obtain equations~\eqref{7} and~\eqref{7b}, which confirm our initial observation, namely, that in the high density limit, $P_{fc}$ is dominated by nodes which are situated in ``\textit{hard to connect}" regions, characterized by their solid angle. Using \eqref{H} we then instructively worked through three example domains (circle, square and circular sector) in order to reach our second main result, a general formula for $P_{fc}$ (equation~\eqref{gen1}). The closed and simple format of \eqref{gen1} emphasizes the logical decomposition of the domain into objects of different full connectivity importance, according to their codimension. Finally, a comparison of our theoretical predictions with direct numerical simulations, in a variety of settings, demonstrated the inaccuracy of conventional percolation models while emphasizing the benefits of including boundary effects.

In summary, full connectivity at high node densities is dominated by local geometric boundary effects such as corners, edges and faces. These contributions have universal properties, distinct but complementary to those of previous percolation approaches, which are typically limited to bulk effects or use a scaling that removes boundary terms (see \cite{Mao11}).
As a result, our findings bring forward a radically different understanding of connectivity in confined geometries while also providing useful formulas for the probability of full connectivity, crucial for the design of reliable wireless mesh networks \cite{Haenggi09}. Moreover, sufficient quantitative detail is available for analyzing and determining system parameters in order to mitigate boundary effects (see \cite{CDG11}). Conversely, boundary effects can be harnessed to avoid full connectivity. Such approaches can be useful in physically important models such as the spread of forest fires \cite{Pueyo10}, epidemics \cite{Danon11}, or mobile phone viruses \cite{Wang09}, given details of a specific model for connectivity. Finally, one may also consider the inverse problem of ``connecting the shape of a drum'' \cite{Kac66} to characterize unknown domains containing random networks. More generally, boundary effects are of major interest in the statistical mechanics of systems with very small structures such as in conducting carbon nanotubes \cite{Kyrylyuk11} or very large and highly connected social and financial networks \cite{Palla07,Parshani11} where the geometry is likely to be dynamic.

\section*{Acknowledgments}
The authors would like to thank the directors of the Toshiba Telecommunications Research Laboratory for their support.

\end{document}